\documentclass[11pt,a4paper]{article}
\pdfoutput=1

\usepackage{jheppub}
\usepackage{bbold}

\title{Strong lensing constraints on bimetric massive gravity}
\author[a]{Jonas Enander}
\author[a]{Edvard M\"ortsell}
\affiliation[a]{Oskar Klein Center, Stockholm University,\\Albanova University Center\\ 106 91 Stockholm, Sweden}
\emailAdd{enander@fysik.su.se}
\emailAdd{edvard@fysik.su.se}
\abstract{We derive dynamical and gravitational lensing properties of local sources in the Hassan-Rosen bimetric gravity theory. Observations of elliptical galaxies rule out values of the effective length-scale of the theory, in units of the Hubble radius, in the interval $10^{-6}\lesssim \lambda_g/r_H\lesssim 10^{-3}$, unless the proportionality constant between the metrics at the background level is far from unity, in which case general relativity is effectively restored for local sources. In order to have background solutions resembling the concordance cosmological model, without fine-tuning of the parameters of the model, we are restricted to the upper interval, or $\lambda_g/r_H\sim 1$, for which the Vainshtein mechanism is expected to restore general relativity for local sources. Except for a limited range of parameter values, the Hassan-Rosen theory is thus consistent with the observed lensing and dynamical properties of elliptical galaxies.}
\begin{document}

\maketitle
 
\section{Introduction}

The recently formulated Hassan-Rosen (HR) theory \cite{HassanRosen2012b, HassanRosen2012c}, which is a ghost-free bimetric theory of gravity, has a rich phenomenology. The theory has been shown to be able to yield background solutions indistinguishable from  a $\Lambda$CDM universe on the background level, even when no explicit cosmological constant or vacuum energy is included in the model \cite{DeFelice2013, Volkov:2013roa, vonStrauss:2011mq, Maeda:2013bha, Volkov:2011an, Volkov:2012zb, Capozziello:2012re, Akrami:2012vf, Akrami:2013ffa}. It is also possible to generate accelerating models that deviate from that of a pure cosmological constant universe. These degeneracies on the background level are broken when studying structure formation in the linear regime, although explicit constraints on the parameters of the model arising from this fact are yet to be obtained \cite{Berg:2012kn, Comelli:2012db, Sakakihara:2012iq, Khosravi:2012rk, Kuhnel:2012gh}. 

The HR theory was developed as an extension of the de Rham-Gabadadze-Tolley (dRGT) theory \cite{deRhamGabadadze2010, dRGT2010, HassanRosen2011}, in conjunction of proving the ghost-free nature of the latter theory \cite{HassanRosen2012a, HassanRosenSchmidtMay2012}. The original motivation of constructing the dRGT theory, in turn, went back all the way to Fierz's and Pauli's original investigations into the formulation of a consistent theory of massive spin-2 fields \cite{Fierz1939, FierzPauli1939}. Formulations in the direction of a fully non-linear, ghost-free theory had previously been performed in e.g. \cite{IshamSalamStrathdee1971, SalamStrathdee1976, ChamseddineSalamStrathdee1978, ArkaniHamed2003, Creminelli2005, Groot2007}.

The idea of introducing a massive spin-2 field to general relativity is intriguing based on the, by now, well-established late-time acceleration of the expansion rate of the universe. As stated above, for a small graviton mass, the HR theory can address the issue of the observed acceleration. In this paper, we investigate whether the theory is compatible with, and/or how the theory can be constrained from observations on galactic scales and below. With spherically symmetric solutions in the HR theory (previously studied in \cite{Comelli:2011wq, Volkov:2012wp, Volkov:2013roa, Babichev:2013una}), we can use observations of galactic velocity dispersions and gravitational lensing angles to constrain the parameters of the theory. 

In sec.~\ref{sec:HR}, the basic aspects of HR theory used in this paper are summarized. In sec.~\ref{sec:sss}, spherically symmetric solutions weak field solutions with and without sources are presented, with the special case of point mass sources given in sec.~\ref{sec:pm}. The effect of including higher order terms, i.e. the Vainshtein mechanism, is discussed in sec.~\ref{sec:vr}. In sec.~\ref{sec:la}, \ref{sec:data} and \ref{sec:results}, we present the method, observational data and results in terms of constraints on the model parameter values. We conclude in sec.~\ref{sec:conclusions}.    

\section{Hassan-Rosen bimetric massive gravity}\label{sec:HR}

The Hassan-Rosen formulation of bimetric massive gravity is given by the following action:
\begin{equation}
S=\int d^{4}x\left[\frac{M_{g}^{2}}{2}\sqrt{-g}R_{g}+\frac{M_{f}^{2}}{2}\sqrt{-f}R_{f}-2m^{2}M_{g}^{2}\sqrt{-g}\sum_{n=0}^{4}\beta_{n}e_{n}\left(\sqrt{g^{-1}f}\right)+\sqrt{-g}\mathcal{L}_{m}\left(g,\Phi\right)\right].
\end{equation}
Here $\mathcal{L}_{m}$ is the matter Lagrangian coupled to $g_{\mu\nu}$. In principle, it is possible to also add a different matter Lagrangian coupled to $f_{\mu\nu}$, but in this paper we choose not to do so. The functions $e_{n}$ (with matrix arguments) are not needed in this paper, but can be found in e.g. \cite{vonStrauss:2011mq}.
Varying the equations of motion with respect to $g_{\mu\nu}$ and $f_{\mu\nu}$ gives the following equations of motion:
\begin{equation}
R_{\mu\nu}\left(g\right)-\frac{1}{2}g_{\mu\nu}R\left(g\right)+m^{2}\sum_{n=0}^{3}\left(-1\right)^{n}\beta_{n}g_{\mu\lambda}Y_{\left(n\right)\nu}^{\lambda}\left(\sqrt{g^{-1}f}\right)=\frac{1}{M_{g}^{2}}T_{\mu\nu},
\label{eq:eomg}
\end{equation}
\begin{equation}
R_{\mu\nu}\left(f\right)-\frac{1}{2}f_{\mu\nu}R\left(f\right)+m^{2}\frac{M_{g}^{2}}{M_{f}^{2}}\sum_{n=0}^{3}\left(-1\right)^{n}\beta_{4-n}f_{\mu\lambda}Y_{\left(n\right)\nu}^{\lambda}\left(\sqrt{f^{-1}g}\right)=0.
\label{eq:eomf}
\end{equation}
Here the matrices $Y_{(n)}$ are given by
\begin{equation}
Y_{\left(0\right)}\left(\mathbb{X}\right)=\mathbb{1},\qquad Y_{\left(1\right)}\left(\mathbb{X}\right)=\mathbb{X}-\mathbb{1}\cdot e_{1}\left(\mathbb{X}\right),
\end{equation}
\begin{equation}
Y_{\left(2\right)}=\mathbb{X}^{2}-\mathbb{X}\cdot e_{1}\left(\mathbb{X}\right)+\mathbb{1}\cdot e_{2}\left(\mathbb{X}\right),
\end{equation}
\begin{equation}
Y_{\left(3\right)}=\mathbb{X}^{3}-\mathbb{X}^{2}\cdot e_{1}\left(\mathbb{X}\right)+\mathbb{X}\cdot e_{2}\left(\mathbb{X}\right)-\mathbb{1}\cdot e_{3}\left(\mathbb{X}\right).
\end{equation}
Taking the divergence, with respect to the $g$-metric, of eq.~(\ref{eq:eomg}), and assuming source conservation, gives the following constraint:
\begin{equation}
\nabla^{\mu}\sum_{n=0}^{3}\left(-1\right)^{n}\beta_{n}\left[g_{\mu\lambda}Y_{\left(n\right)\nu}^{\lambda}\left(\sqrt{g^{-1}f}\right)+g_{\nu\lambda}Y_{\left(n\right)\mu}^{\lambda}\left(\sqrt{g^{-1}f}\right)\right]=0.
\label{eq:gconstraint}
\end{equation}
It can be shown that this constraint is equivalent to the constraint given by taking the divergence with respect to the $f$-metric, of eq.~(\ref{eq:eomf}).

By doing the constant rescalings
\begin{equation}
f_{\mu\nu}\rightarrow\frac{M_{g}^{2}}{M_{f}^{2}}f_{\mu\nu},\qquad\beta_{n}\rightarrow\left(\frac{M_{f}}{M_{g}}\right)^{n}\beta_{n},
\end{equation}
the equations of motion for $f_{\mu\nu}$ become
\begin{equation}
R_{\mu\nu}\left(f\right)-\frac{1}{2}f_{\mu\nu}R\left(f\right)+\frac{m^{2}}{2}\sum_{n=0}^{3}\left(-1\right)^{n}\beta_{4-n}\left[f_{\mu\lambda}Y_{\left(n\right)\nu}^{\lambda}\left(\sqrt{f^{-1}g}\right)+f_{\nu\lambda}Y_{\left(n\right)\mu}^{\lambda}\left(\sqrt{f^{-1}g}\right)\right]=0.
\label{eq:eomf2}
\end{equation}
The ratio $M_{g}/M_{f}$ therefore drops out of the equations of motion. This is a reflection of the fact that we have not coupled $f_{\mu\nu}$ to any gravitational sources. The HR theory thus has five free parameters $\beta_i$, where $i=[0,\ldots,4]$ (remembering that $m$ just multiplies all the $\beta_i$:s to get the correct dimensionality). Two of these, $\beta_{0}$ and $\beta_{4}$ correspond, on the level of the Lagrangian, to a cosmological constant for the $g$- and $f$- sector, respectively. On the level of the equations of motion, however, it will be certain combinations of all the $\beta_i$:s that contribute to an effective cosmological constant.

\section{Spherically symmetric solutions}\label{sec:sss}

Spherically symmetric solutions in the HR theory have previously been studied in \cite{Comelli:2011wq, Volkov:2012wp, Volkov:2013roa, Babichev:2013una}. Because of the absence of an equivalent of Birkhoff's theorem in the HR theory, there does not exist a unique solution for a spherically symmetric static spacetime. The studied solutions fall into two broad classes. With $g_{\mu\nu}$ diagonal, the most general form for $f_{\mu\nu}$, after gauge fixing, contains an off-diagonal element $f_{rt}$. In the case of non-zero $f_{rt}$, \cite{Comelli:2011wq} gave the complete analytical solutions. It turns out to that $g_{\mu\nu}$ in this case is completely degenerate with the standard Schwarzschild-de Sitter (or Kottler) metric. For the ansatz $f_{rt}=0$, the equations of motion turn out to be highly involved. \cite{Comelli:2011wq} wrote down the linear solution, whereas \cite{Volkov:2012wp} did an exhaustive numerical study of the solution. The main result of \cite{Volkov:2012wp} is that, for the diagonal ansatz, there is "a whole zoo of new black holes with massive degrees of freedom excited." 

In this paper, we rederive the linear solution provided by \cite{Comelli:2011wq} but in isotropic form, making the solutions more accessible for a gravitational lensing analysis. 
We also include second order terms to compute the size of the Vainshtein radius. This is to make sure that a linear analysis is valid in the region accessible for phenomenological study. 
Furthermore, the inclusion of matter sources allows us to observationally constrain the parameters of the theory. As our ans{\"a}tze, we use the following diagonal forms for $g_{\mu\nu}$ and $f_{\mu\nu}$:
\begin{equation}
ds_{g}^{2}=-V^{2}dt^{2}+W^{2}\left(dr^{2}+r^{2}d\Omega^{2}\right),
\end{equation}
\begin{equation}
ds_{f}^{2}=-A^{2}dt^{2}+B^{2}dr^{2}+C^{2}r^{2}d\Omega^{2}.
\end{equation}
We perturb the metric around flat space (where the  background metric is flat, i.e. $\bar{g}_{\mu\nu}=\eta_{\mu\nu}$ and $\bar{f}_{\mu\nu}=c^2\eta_{\mu\nu}$) in the following way:
\begin{equation}
V\simeq1+\delta V,\quad W\simeq1+\delta W,
\end{equation}
\begin{equation}
A\simeq c\left(1+\delta A\right),\quad B\simeq c\left(1+\delta B\right),\quad C\simeq c\left(1+\delta C\right).
\end{equation}
Since $g_{\mu\nu}$ is put on isotropic form, we can identify $\delta V=\Phi$ and $\delta W=\Psi$, where $\Phi$ is the gravitational potential and $\Psi$ is the spatial curvature for scalar perturbations in the Newtonian gauge.
For flat space to be a valid background solution, we must impose
\begin{equation}
\beta_{0}+3\beta_{1}c+3\beta_{2}c^{2}+\beta_{3}c^{3}=0,
\end{equation}
\begin{equation}
\beta_{4}c^{3}+3\beta_{3}c^{2}+3\beta_{2}c+\beta_{1}=0,
\end{equation}
in order to remove cosmological constant contributions in the $g$- and $f$-sector. We note that this corresponds to the case where the background expansion is pure GR (but it is still possible, however, that perturbations around the background deviate from GR). Notice that the $\beta_{i}$:s are parameter in the Lagrangian, whereas $c$ is a parameter of the background solution. To first order, the solutions to the equations of motion given in eq.~(\ref{eq:eomg}) and (\ref{eq:eomf2}) in vacuum are
\begin{equation}
\delta V=-\frac{GM_{1}}{r}-\frac{c^{2}GM_{2}}{r}e^{-m_{g}r},
\end{equation}
\begin{equation}
\delta W=\frac{GM_{1}}{r}+\frac{c^{2}GM_{2}}{2r}e^{-m_{g}r},
\end{equation}
\begin{equation}
\delta A=-\frac{GM_{1}}{r}+\frac{GM_{2}}{r}e^{-m_{g}r},
\end{equation}
\begin{equation}
\delta B=\frac{GM_{1}}{r}+\frac{GM_{2}\left[2\left(1+c^{2}\right)\left(1+m_{g}r\right)+c^{2}m_{g}^{2}r^{2}\right]}{2m_{g}^{2}r^{3}}e^{-m_{g}r},
\end{equation}
\begin{equation}
\delta C=\frac{GM_{1}}{r}-\frac{GM_{2}\left[\left(1+c^{2}\right)\left(1+m_{g}r\right)+m_{g}^{2}r^{2}\right]}{2m_{g}^{2}r^{3}}e^{-m_{g}r},
\end{equation}
where
\begin{equation}\label{eq:mg}
m_{g}^{2}\equiv m^{2}\left(c+c^{-1}\right)\left(\beta_{1}+2\beta_{2}c+\beta_{3}c^{2}\right).
\end{equation}
Here $M_{1}$ and $M_{2}$ are arbitrary integrations constants. The second order solutions are given in appendix \ref{app:sol2}.

If we introduce a pressureless source, for which $T^{0}_{0}=-\rho$, and define
\begin{equation}
\Theta_{ml}\equiv c^{2}\delta A+\delta V,\qquad\Theta_{m}\equiv\delta A-\delta V,
\end{equation}
where $ml$ and $m$ stands for massless and massive, respectively, we get the following two source equations for $\Theta_{ml}$ and $\Theta_{m}$ (for more details on the identification of the massless and massive modes in the HR theory see \cite{Hassan:2012wr}):
\begin{equation}\label{eq:thetaml}
\nabla^{2}\Theta_{ml}=4\pi G\rho,
\end{equation}
\begin{equation}
\left(\nabla^{2}-m^{2}_{g}\right)\Theta_{m}=-\frac{16\pi G}{3}\rho.
\end{equation}
Inverting these equations gives
\begin{equation}\label{eq:thetaml1}
\Theta_{ml}(\mathbf{r}) =-G\int d^{3}\mathbf{r}^{\prime}\frac{\rho\left(\mathbf{r}^{\prime}\right)}{\left|\mathbf{r}-\mathbf{r}^{\prime}\right|}+\mbox{b.c.},
\end{equation}
\begin{equation}\label{eq:thetam1}
\Theta_{m}(\mathbf{r}) =\frac{4G}{3}\int d^{3}\mathbf{r}^{\prime}\frac{\rho\left(\mathbf{r}^{\prime}\right)}{\left|\mathbf{r}-\mathbf{r}^{\prime}\right|}e^{-m_{g}\left|\mathbf{r}-\mathbf{r}^{\prime}\right|}+\mbox{b.c.}.
\end{equation}
Solving for $\delta V$, and putting the boundary terms (b.c.) to zero, then gives
\begin{equation}
\delta V(\mathbf{r}) =\frac{\Theta_{ml}-c^{2}\Theta_{m}}{1+c^{2}}= -\frac{G}{1+c^{2}}\int d^{3}\mathbf{r}^{\prime}\frac{\rho\left(\mathbf{r}^{\prime}\right)}{\left|\mathbf{r}-\mathbf{r}^{\prime}\right|}\left(1+\frac{4c^{2}}{3}e^{-m_{g}\left|\mathbf{r}-\mathbf{r}^{\prime}\right|}\right).
\label{eq:deltaV}
\end{equation}
Notice that after introducing a source we do not have two independent integration constants; $\delta V$ is completely determined by the source and the model parameters $m_g$ and $c$. Since the normalization of $\rho$ (or equivalently $M_1$ and $M_2$ in terms of the vacuum solutions) is arbitrary, in the following, we employ a constant rescaling of Newton's constant
\begin{equation}
\frac{G}{1+c^2}\rightarrow G.
\end{equation}
Note that a large value of the proportionality constant $c$ in this sense could perhaps be related to the small observed value of $G$.

Since the HR theory represents a generalization of GR, it is natural to ask the question of whether the theory is capable of explaining the rotation curves of spiral galaxies without introducing a dark matter halo component. 
However, including only first order perturbations in the spherically symmetric solutions, this is not possible since the observed rotation curves generally are flatter at large radii than what can be obtained using the baryonic matter distribution only. The inclusion of an additional Yukawa term will have the opposite effect of increasing the slope as the Yukawa term decays after which the asymptotic behaviour equals the standard Newtonian form. The inclusion of the Yukawa term thus pushes the peak of the rotational velocity toward lower radii, as compared to the case of a purely Newtonian rotation curve. However, this conclusion may be altered when including higher order terms, that is the Vainshtein mechanism, in the solutions \cite{rotcurve}. Also note that eq.~(\ref{eq:thetaml}--\ref{eq:deltaV})  were computed in the weak-field limit, whereas for the interesting case of  $\lambda_{g}\sim r_{H}$ higher-order terms have to be included. 

\section{Point mass source solutions}\label{sec:pm}

Introducing a point mass source with mass $M$, i.e. putting $\rho=M\delta^{(3)} (r)$ in eq.~(\ref{eq:thetaml1}--\ref{eq:thetam1}), gives the following first order solutions:
\begin{eqnarray}
\Phi & = & -\frac{GM}{r}(1+\frac{4c^{2}}{3}e^{-m_{g}r}),\\
\Psi & = & -\frac{GM}{r}(1+\frac{2c^{2}}{3}e^{-m_{g}r}),\\
\varphi & = & -\frac{GM}{r}(1+c^{2}e^{-m_{g}r}),
\end{eqnarray}
where $\Phi$ is the gravitational potential, $\Psi$ the spatial curvature and $\varphi=(\Phi+\Psi)/2$ is the effective gravitational potential felt by massless particles. We can decompose the potentials as $\Phi=\Phi_{GR}+\Phi_{Y}$, $\Psi=\Psi_{GR}+\Psi_{Y}$ and $\varphi=\varphi_{GR}+\varphi_{Y}$, where the subscript $GR$ denotes the general relativity value of $-M/r$ and $Y$ the Yukawa terms given by
\begin{eqnarray}
\Phi_{Y} & = & -\frac{4c^{2}GM}{3r}e^{-m_{g}r},\\
\Psi_{Y} & = & -\frac{2c^{2}GM}{3r}e^{-m_{g}r},\\
\varphi_{Y} & = & -\frac{c^{2}GM}{r}e^{-m_{g}r}.
\end{eqnarray}
We note that $M$ is the mass we would measure for a point mass at infinite distance. As evident from eq.~(\ref{eq:mg}), $m_{g}$ and $c$ are not independent parameters. In fact, the Yukawa terms will approach zero both as $c\rightarrow0$ and $c\rightarrow\infty$. These asymptotes are independent of the values of the $\beta$'s.
To understand the behaviour in between, we may assume that $\beta_{1}\sim\beta_{2}\sim\beta_{3}=\beta$ and thus
\begin{equation}
m_{g}^{2}=\beta m^{2}\frac{(c^{2}+1)(c+1)^{2}}{c}\equiv m_{b}^{2}\frac{(c^{2}+1)(c+1)^{2}}{c},
\end{equation}
where we have defined $m_{b}^{2}\equiv\beta m^{2}$. That is, the only way to have sizable modifications of the potentials on galactic scales from the Yukawa terms when $c$ is not too far from unity, is to have $\sqrt{\beta}m$ of the inverse order of galactic scales or smaller. 

In the limit $m_{g}r\rightarrow0$, the ratio of the gravitational potentials felt by massive and massless particles is given by 
\begin{equation}
\frac{\Phi}{\varphi}=\frac{1+(4c^2)/3}{1+c^2}.
\end{equation}
Note the similarity to the vDVZ-discontinuity factor of $4/3$ in linear massive gravity. 
However, it is expected that this discrepancy between massive and massless particles as $m_{g}r\rightarrow0$ will be removed as we include higher order terms, see sec.~\ref{sec:vr}.

\section{The Vainshtein radius}\label{sec:vr}

In 1972 Vainshtein observed that the formulation of massive gravity given at the time exhibited a radius that signals the breakdown of the linear expansion around a source with mass $M$ (\cite{Vainshtein1972}, see also \cite{Babichev:2013usa} for a recent review). This radius was latter called the Vainshtein radius, and is given by
\begin{equation}
r_{V}\equiv \left(\frac{GM}{m^{4}}\right)^{1/5}\sim\left(r_S\lambda^4\right)^{1/5},
\end{equation}
where $r_S$ is the Schwarzschild radius of the source and $\lambda$ the wavelength associated to the massive graviton, i.e. $\lambda=m^{-1}$.
Within this radius, higher-order corrections to the expansion of the metric in powers of $GM$ have to be taken into account. 
Since $r_V$ is an intermediate scale between the gravitational scale of the source and the Compton wavelength of the graviton, 
for any specific source and graviton mass, one has to make sure to be well outside $r_V$ for the linear expansion is to be valid. The Vainshtein radius is derived for $r\ll\lambda$, which will always hold for local sources when $\lambda$ is of the order of the Hubble radius. For $r\simeq\lambda$, however, the Vainshtein radius is not applicable, and one has to check numerically that the second order solution does not dominate over the first order solution in the region of interest. 

In order to identify the Vainshtein radius for spherically symmetric sources in the HR theory,  we have solved the diagonal ansatz to second order.\footnote{An extensive study of the Vainshtein mechanism in the HR theory appeared in \cite{Babichev:2013pfa} after this paper was written.} The solution is given in appendix \ref{app:sol2}. 
At second order, two new effective parameters occur, namely
\begin{equation}
m_{1}^{2}\equiv m^{2}\left(\beta_{1}+\beta_{2}c\right),\quad m_{2}^{2}\equiv m^{2}\left(\beta_{2}c+\beta_{3}c^{2}\right).
\end{equation}
These are related to $m_{g}^{2}$ through
\begin{equation}
m_{g}^{2}=\left(c^{-1}+c\right)\left(m_{1}^{2}+m_{2}^{2}\right).
\end{equation}
In the $m_{g}r\ll1$ limit, the full second order solution then has a dominant term from which one can read off $r_V$ as the radius where second order terms start dominating over first order terms, 
\begin{equation}\label{eq:rv}
r_{V}\equiv\left[\frac{GM\left(1+c^{2}\right)^{3}m_{2}^{2}}{c^{3}m_{g}^{4}}\right]^{1/3}.
\end{equation}
This holds for all fields expect $F$, for which we instead have
\begin{equation}
r_{V}^{A}\equiv\left[\frac{GM\left(1+c^{2}\right)^{3}\left(7m_{1}^{2}+8m_{2}^{2}\right)}{c^{3}m_{g}^{4}}\right]^{1/3}.
\end{equation}
This means that it is not possible to decrease $r_V$ to smaller values by letting $m_{2}^{2}\rightarrow 0$ (i.e. looking at the other terms in the second order solution). 
Putting $m_{1}^{2}=0$, so that
\begin{equation}
m_{g}^{2}=\frac{1+c^{2}}{c}m_{2}^{2},
\end{equation}
gives the Vainshtein radius, common to all fields, as
\begin{equation}
r_{V}=\left[\frac{GMc^2\left(1+c^{-2}\right)^{2}}{m_{g}^{2}}\right]^{1/3}.
\end{equation}
Numerically, this is given by (for $c=1$)
\begin{equation}
r_{V}\simeq 0.17\left[\left(\frac{M}{M_\odot}\right)\left(\frac{\lambda_g}{r_H}\right)^2\right]^{1/3}\mbox{kpc}\simeq 3.4\cdot 10^{-8}\left[\left(\frac{M}{M_\odot}\right)\left(\frac{\lambda_g}{r_H}\right)^2\right]^{1/3}r_H,
\end{equation}
where $\lambda_g\equiv m_g^{-1}$ and $r_H=H_0^{-1}$ is the Hubble radius. For a galactic mass scale of $M\sim 10^{11}M_{\odot}$ and $\lambda_g\sim r_H\sim 5\cdot 10^6\,{\rm kpc}$, we obtain 
$r_V\sim 800\,{\rm kpc}$, i.e., more than a factor of 100 larger than the radius probed by the observations used in this paper. We also note that for the Sun as the gravitational source, the Vainshtein radius is larger than 1 AU as long as $\lambda_g\gtrsim 5\cdot 10^{-12}\,r_H$.  
In the following, we have assumed that eq.~\ref{eq:rv} is a fair approximation for $r_V$ even when $r\lesssim\lambda_g$. That is, that we  are well outside the Vainshtein radius when constraining the Yukawa decay of the potentials, making it possible to constrain the parameters $m_{g}$ and $c$ using the linear approximation. 

\section{Lensing analysis}\label{sec:la}

Since massive and massless particles experience different forces in a gravitational field in bimetric theories, we can constrain such theories if we have access to systems where the gravitational field, or mass, is probed by both massive particles and photons. One such example is the Sun, which we will return to later. On larger scales, galaxies and galaxy clusters, where we have both dynamical and lensing data, are obvious candidates. In this paper, we will make use of elliptical galaxies for which we have measurements of both the velocity dispersion and the gravitational lensing deflection angle. 
In doing this, we will to large extent apply the same methodology and data as in \cite{2010ApJ...708..750S} and \cite{2011arXiv1111.5961S}. 
Basically, the method amounts to investigating for which parameter ranges of the theory the galaxy masses as inferred from massive particles (velocity dispersions) and massless particles (lensing angle) are consistent.

The velocity dispersion in elliptical galaxies can be derived from the equations of stellar hydrodynamics:
\begin{equation}
\frac{d}{dr}(\nu\sigma_{r}^{2})+\frac{2\zeta}{r}\nu\sigma_{t}^{2}=-\nu\frac{d\Phi}{dr}\equiv-\nu\Phi^{\prime},
\end{equation}
where $\sigma_{t}$ and $\sigma_{r}$ are the velocity dispersions in the tangential and radial direction, respectively, $\zeta=1-(\sigma_{t}/\sigma_{r})^{2}$ is the velocity anisotropy, $\nu$ is the density of velocity dispersion tracers (in this case the luminous matter) and $\Phi$ is the total gravitational potential. The prime indicates differentiation with respect to $r$. Assuming that $\zeta$ is constant, we can write
\begin{equation}\label{eq:sigmar2}
\sigma_{r}^{2}(r)=\frac{1}{\nu r^{2\zeta}}\intop_{r}^{\infty}\nu r^{2\zeta}\Phi^{\prime}dr.
\end{equation}
Note that the integral is from $r$ to $\infty$, the reason being that it is normalized such that the velocity dispersion approaches zero asymptotically. The actual observed velocity dispersion, given by the single number $\sigma_{\star}^{2}$, is then given by a line-of-sight luminosity weighted average over the effective spectroscopic aperture of the observations. To compute the velocity dispersion, we need $\nu(r)$, $\zeta$ and the radial derivative of the gravitational potential, given by the total density distribution $\rho(r)$. In the following, we assume that both the luminous and total matter distribution can be written as power laws 
\begin{eqnarray}
\nu(r) & = & \nu_{0}\left(\frac{r}{r_{0}}\right)^{-\delta}\\
\rho(r) & = & \rho_{0}\left(\frac{r}{r_{0}}\right)^{-\gamma}.
\end{eqnarray}
In appendix~\ref{sec:Velocity-dispersions}, general expressions to derive observed velocity dispersions in the HR theory are outlined.

The deflection angle of photons passing through a gravitational field is given by
\begin{equation}
\hat{\alpha}=2\intop_{-\infty}^{\infty}\nabla_{\bot}\varphi dl,\label{eq:alpha}
\end{equation}
where the integral to excellent approximation can be calculated over the undeflected path of the photon, and the derivative is with respect to the direction perpendicular to the direction of the photon. Now, what is actually observed is the angular image separation between multiple images of a background source.  
If the observer, lens and source are perfectly aligned, the appropriately scaled deflection angle is given by half the observed angular image separation, the so called Einstein angle of the system. It can be shown that this is an excellent approximation even in cases when the source does not lie directly behind the lens, and we can thus use the observed image separation to estimate the deflection angle.  In appendix~\ref{sec:Gravitational-lensing}, general expressions to derive the deflection angles in HR theory are derived. 

In practice, given values for the model parameters $m_g$ and $c$, we can now use the observed deflection angles to normalize the mass density profile, or $\rho_{0}r_{0}^{\gamma}$, of each galaxy which is then used to predict a value of the velocity dispersion that can be compared to the observed value. 
The analysis is complicated by the fact that the force experienced by massive and massless particles are not fully determined by the mass inside the radius at hand. However, these complications can be overcome by using the approximations outlined in appendix~\ref{sec:Velocity-dispersions} and \ref{sec:Gravitational-lensing}.

Note that the method of comparing gravitational deflection angles with the dynamics of massive particles makes us very insensitive to the assumed matter distribution of the galaxies, specifically since the deflected photons and the velocity dispersion tracers effectively probes similar galactic radii of $r_h\sim 10^{-6} r_H\sim 5\,{\rm kpc}$. We also note that given prior knowledge on the normalization of the individual mass density profiles, we could in principle use the observed velocity dispersions and gravitational lensing angles individually to constrain the parameters of the model. For example, for a given mass distribution, we expect the observed velocity dispersion in HR theory to be larger than in GR and for a given observed velocity dispersion, the mass-to-light ratio required in HR theory to be smaller than in GR. Such an analysis is left for future work.  

\section{Data}\label{sec:data}

In this paper, we make use of the strong gravitational lens sample observed with the Hubble Space Telescope Advanced Camera for Surveys by the Sloan Lens ACS (SLACS) Survey \cite{2008ApJ...682..964B}. The full sample consists of 131 strong lens candidates out of which we make use of a sub-sample of 53 systems with elliptical lens galaxies, well fitted by singular isothermal ellipsoidal lens models, and having reliable velocity dispersion measurements\footnote{A compilation of the data employed in this paper for the subsample of 53 systems can be found at \tt{http://www.physto.se/\textasciitilde edvard/slacsl.html}.}. We use the velocity dispersions as measured from Sloan Digital Sky Survey (SDSS) spectroscopy over an effective spectroscopic aperture of $1.4$ arcsec and the Einstein angle as measured from ACS imaging data. From ACS images, we also use the effective radii of the lens galaxies to individually estimate the luminosity profile power-law index $\delta$ of the lensing galaxies by comparing the total luminosity to the luminosity within half the effective radii. To the approximately $7\,\%$ velocity dispersion fractional errors quoted in \cite{2008ApJ...682..964B}, we add an additional $5\,\%$ to take into account possible deviations from the singular isothermal mass profile \cite{2008A&A...477..397G}. We assume a $2\,\%$ error on the measured image separations. The slope
and anisotropy of the lensing galaxies are being individually marginalized over, using prior probabilities of $\gamma=2.00\pm0.08$ and $\zeta=0.13\pm0.13$ ($68\,\%$ confidence level) \cite{2010ApJ...708..750S}. 

\section{Results}\label{sec:results}

\begin{figure}
\begin{centering}
\includegraphics[scale=0.4]{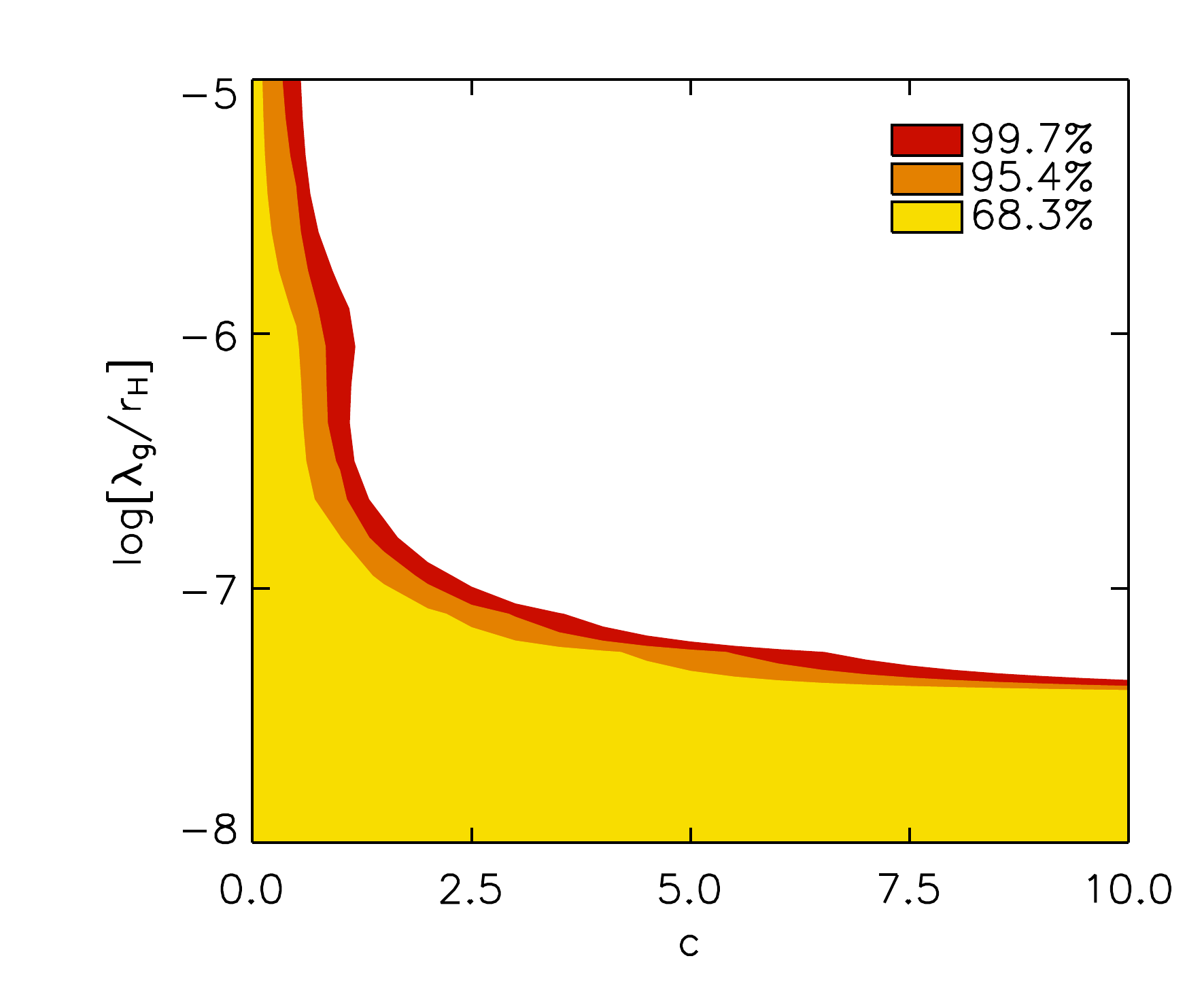}
\includegraphics[scale=0.4]{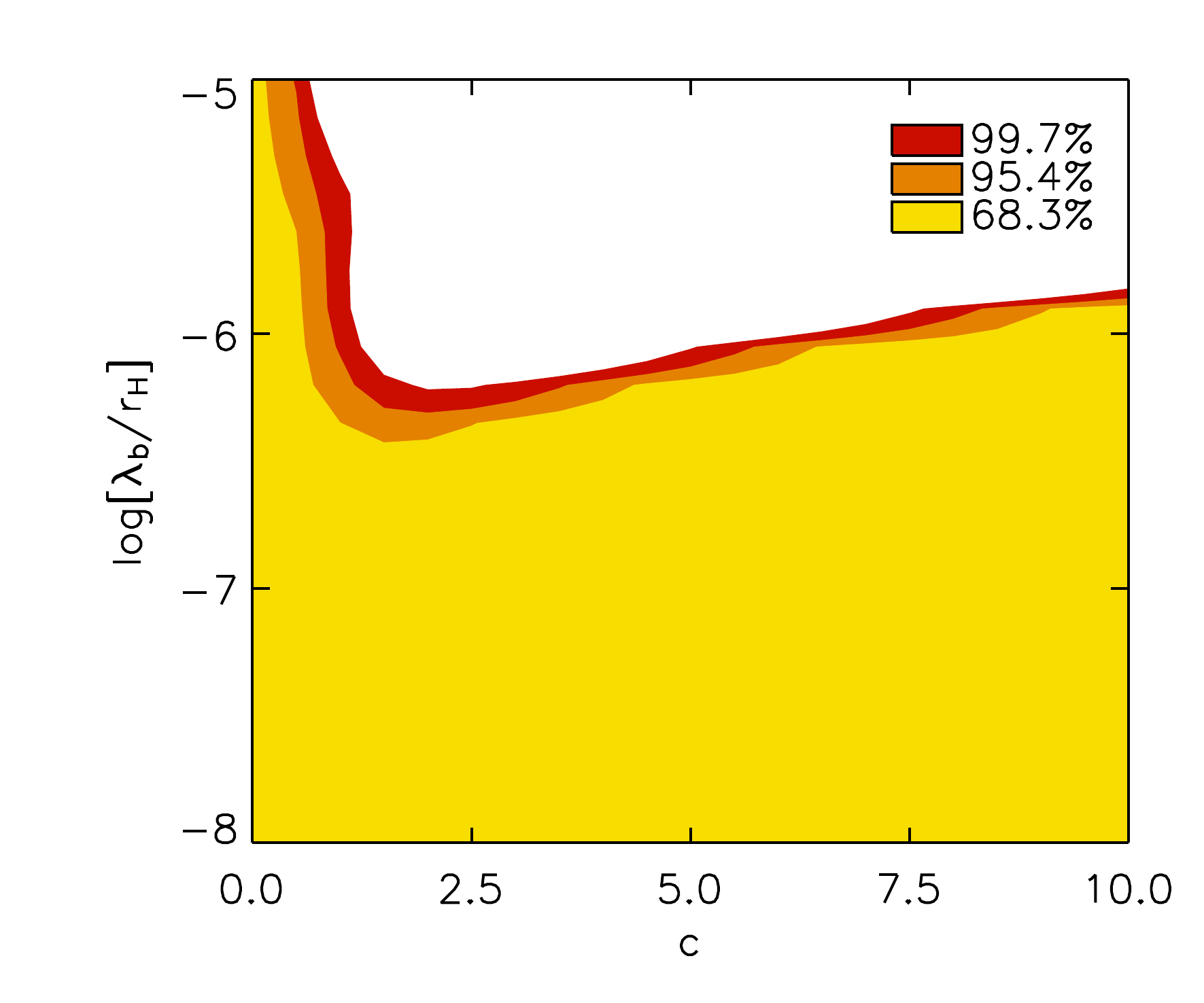}
\par\end{centering}
\caption{\label{fig:mgc2}{\em Left panel:} Observational limits on $\lambda_{g}$ in units of the Hubble radius $r_H=H_0^{-1}$ and $c$. For $c\sim 1$, $\lambda_g\lesssim 10^{-6}\,r_H\sim 5\,{\rm kpc}$. 
{\em Right panel:} Observational limits on $\lambda_b=m_b^{-1}=\lambda_g/\sqrt{\beta}$  in units of the Hubble radius $r_H=H_0^{-1}$. For $c\sim 1$, $\lambda_b\lesssim 10^{-6.3}\sim 2.5\,{\rm kpc}$. As $c\rightarrow0$ and $c\rightarrow\infty$, GR is recovered and $\lambda_b$ becomes unconstrained.}
\end{figure}

Using the method and data described above, we are able to constrain the model parameters $ \lambda_g=m_{g}^{-1}$ and $c$ as depicted in the left panel of fig.~\ref{fig:mgc2}. As anticipated, as $c\rightarrow 0$, $\lambda_g$ becomes unconstrained since GR is recovered in that limit. For $c\sim 1$, data constrains the effective length scale of the theory to be $\lambda_g\lesssim 10^{-6}\,r_H\sim 5\,{\rm kpc}$.

A few comments are in place here: We note that we obtain an upper limit on $\lambda_{g}/r_{H}$ .
The reason for this is that if $\lambda_{g}/r_{H}$ becomes too large, we will have a constant vDVZ-like off-set between the force experienced by massive and massless particles. If we include also non-linear effects in the analysis, we expect the difference between the force experienced by massive and massless particles to be zero at small $r$ or big $\lambda_{g}$, reach a maximum value around the Vainshtein radius and then approach zero again as $r\gg\lambda_{g}$. This would then mean that our data will allow for either large values of $\lambda_{g}\gtrsim 10^{-3}\,r_H$ in which case the galactic scale $r_g\sim 5\,{\rm kpc}$ would be within the Vainshtein radius where GR is restored, or very small values of $\lambda_{g}\lesssim 10^{-6}\,r_H$ where the exponential decay of the Yukawa terms again restores GR. 
This can be compared to the results of \cite{2011arXiv1111.5961S} where a lower limit of $\lambda_{g}/r_{H}\gtrsim0.02$ was obtained for the decoupling limit of the massive gravity model of \cite{2011PhRvD..83j3516D}. 

As noted in sec.~\ref{sec:pm}, $m_{g}$ and $c$ are not independent parameters in terms of the fundamental model parameters. Using the definition $m_{b}^{2}\equiv\beta m^{2}$ where $\beta=\beta_{1}\sim\beta_{2}\sim\beta_{3}$, we can constrain the corresponding length scale $\lambda_{b}=m_b^{-1}=\lambda_g/\sqrt{\beta}$ together with $c$. Results are shown in the right panel of fig.~\ref{fig:mgc2}. As expected, as $c\rightarrow 0$ and $c\rightarrow\infty$, constraints on $\lambda_b$ weakens, but if $c$ is not too far from unity and $\lambda_b$ of the order of galactic scales or larger, we will have sizable contributions from the Yukawa parts of the potentials. For $c\sim 1$, the length scale of the theory is constrained to $\lambda_{b}\lesssim 10^{-6.3}\,r_H\sim 2.5\,{\rm kpc}$, in the linear approximation. Including higher order terms, the Vainshtein mechanism again opens up for the possibility of large values of $\lambda_b$, putting galactic scales within their corresponding Vainshtein radii. 
 
The magnitude of an additional Yukawa term to the GR gravitational potential have been constrained to be very small on scales from our Solar system down to millimeter distances \cite{2005IJMPA..20.2294R,2012PhR...513....1C}. Also, the deflection and time delay of light passing close to the limb of the Sun shows that the  gravitational potential, $\Phi$, and the spatial curvature $\Psi$ are equal up to a fractional difference of $\sim 10^{-5}$ \cite{2003Natur.425..374B}. 
Therefore, unless $\lambda_g$ is in the sub-millimeter range, at Solar system scales ($1\,{\rm AU}\sim 5\cdot 10^{-9}\,{\rm kpc}$), we need to be well within the Vainshtein radius of the Sun for the theory to survive, limiting $\lambda_g\gtrsim 5\cdot 10^{-12}\,r_H\sim 0.025\,{\rm pc}$.

Although we have obtained the spherically symmetric solutions in a background equivalent to GR, we may assume that locally they are useful approximations also in a more general background. To have accelerating cosmological concordance-like solutions, we  need $\lambda_b/r_H\sim1$ \cite{vonStrauss:2011mq,Akrami:2012vf}. For such values, the observational probes employed in this paper are well inside their Vainshtein radii, effectively restoring GR. 

\begin{figure}
\begin{centering}
\includegraphics[scale=0.6]{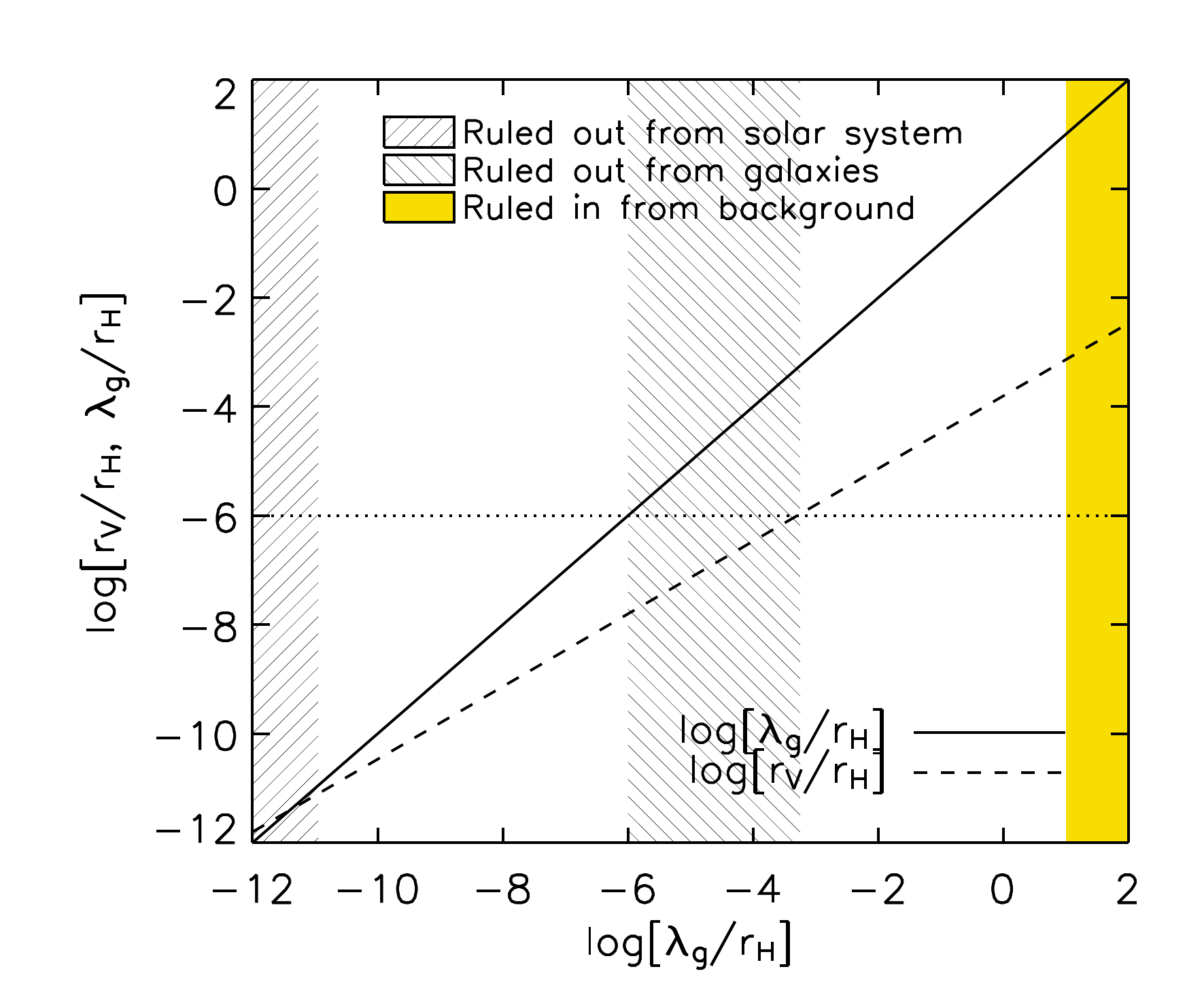}
\par\end{centering}
\caption{\label{fig:lg}Limits on $\lambda_{g}$ including the fact that GR is restored inside the Vainshtein radius and outside the Yukawa length-scale $\lambda_g$. We have assumed $m_1=0$ and $c=1$. Note that for $c$ very different from unity, GR is practically restored at all scales.}
\end{figure}

We can now combine the limits discussed above into fig.~\ref{fig:lg}, where we show the galactic Vainshtein radius in units of $r_H$ (neglecting possible modifications when $r\lesssim\lambda_g$) as a function of the length-scale of the Yukawa decay of spherically symmetric solutions of the bimetric theory. The typical length scale ($r_g\sim 5\,{\rm kpc}$) probed by the velocity dispersion and gravitational lensing observations is indicated by the horizontal dotted line. Going from left to right on the $x$-axis, we make the following observations:
\begin{itemize}
\item Values of $\lambda_g/r_H\lesssim 10^{-11}$ are ruled out from gravity tests on Solar system scales and below.
\item For $10^{-11}\lesssim \lambda_g/r_H\lesssim 10^{-6}$, the scale of the galactic observations, $r_g$, is larger than $\lambda_g$, the Yukawa terms becomes negligible and GR is effectively restored.
\item For $10^{-6}\lesssim \lambda_g/r_H\lesssim 10^{-3}$, $r_g$ is smaller than $\lambda_g$ and the difference of proportionality between the Yukawa terms in the gravitational potential and spatial curvature, invalidates this parameter range when comparing velocity dispersions and lensing deflections.
\item  For $\lambda_g/r_H\gtrsim 10^{-3}$, our observations fall inside the Vainshtein radii of the systems, and the parameter range is ruled in since GR is presumably restored through the Vainshtein mechanism.
\item  Apart from being compatible with observations on galactic scales, values of $\lambda_g/r_H\gtrsim 1$ also have the possibility of providing an explanation of the apparent accelerating expansion of space on cosmological scales. 
\end{itemize}

\section{Conclusions}\label{sec:conclusions}

In this paper we have studied perturbative solutions for a diagonal ansatz for spherically symmetric solutions in the Hassan-Rosen theory. We have compared these solutions with gravitational lensing deflection angles of elliptical galaxies. Using lensing dispersion data we have shown that, for the proportionality constant $c$ not too far from unity, the effective length scale of the theory $\lambda_{g}$ either has to be small enough for the Yukawa term to be negligible an galactic scales, $\lambda_g\lesssim 5\,{\rm kpc}$, or large enough for the radii probed to be within the Vainshtein radii of the galaxies, $\lambda_g\gtrsim 5\,{\rm Mpc}$. Values of $\lambda_g\lesssim 0.025\,{\rm pc}$ are ruled out from observations on Solar system scales and below. We note that if $\lambda_g\sim r_H$, i.e. if the length scale of the theory is close to the Hubble radius, apart from being compatible with data on galactic scales and below due to a presumed Vainshtein radius \cite{Babichev:2013pfa}, the HR theory may also provide a mechanism for the apparent accelerated expansion rate of the Universe. 

\acknowledgments

The authors would like to thank Mikael von Strauss, Fawad Hassan, Angnis Schmidt-May, Bo Sundborg, Florian K\"uhnel and Ariel Goobar for useful discussions. EM acknowledges the Swedish Research Council for financial support.

\appendix

\section{Second order solutions}
\label{app:sol2}

\begin{eqnarray}
\nonumber
\frac{\delta V_{2}}{G^{2}} & = & \frac{M_{1}^{2}}{2r^{2}}+\frac{M_{1}M_{2}}{r^{2}}e^{-m_{g}r}-\frac{M_{2}^{2}\left(1+c^{2}\right)^{2}m_{2}^{2}}{16c^{3}m_{g}^{4}r^{4}}e^{-2m_{g}r}\\ \nonumber
 &  & -\frac{M_{2}^{2}\left(1+c^{2}\right)^{2}m_{2}^{2}}{8c^{3}m_{g}^{3}r^{3}}e^{-2m_{g}r}+\frac{3M_{2}^{2}\left(1+c^{2}\right)\left[\left(-1+4c^{2}\right)m_{1}^{2}+\left(-4+c^{2}\right)m_{2}^{2}\right]}{32c^{3}m_{g}^{2}r^{2}}e^{-2m_{g}r}-\frac{3M_{2}^{2}m_{g}}{8c^{2}r}e^{-2m_{g}r}\\ \nonumber
 &  & +\frac{M_{1}M_{2}m_{g}}{r}e^{-m_{g}r}\log\left(\frac{r}{r_{0}}\right)-\frac{M_{2}^{2}\left(1+c^{2}\right)\left[\left(-25+44c^{2}\right)m_{1}^{2}+\left(25c^{2}-44\right)m_{2}^{2}\right]}{64c^{3}m_{g}r}e^{m_{g}r}\mbox{Ei}\left(-3m_{g}r\right)\\ \nonumber
 &  & -\frac{M_{1}M_{2}m_{g}}{r}e^{m_{g}r}\mbox{Ei}\left(-2m_{g}r\right)-\frac{3M_{2}^{2}m_{g}^{2}}{4c^{2}}\mbox{Ei}\left(-2m_{g}r\right)\\
 &  & +\frac{M_{2}^{2}\left(1+c^{2}\right)\left[\left(-25+44c^{2}\right)m_{1}^{2}+\left(25c^{2}-44\right)m_{2}^{2}\right]}{64c^{3}m_{g}r}e^{-m_{g}r}\mbox{Ei}\left(-m_{g}r\right),
\end{eqnarray}

\begin{eqnarray}
\nonumber
\frac{\delta W_{2}}{G^{2}} & = & \frac{M_{1}^{2}}{4r^{2}}+\frac{M_{1}M_{2}}{8r^{2}}e^{-m_{g}r}+\frac{3M_{1}M_{2}m_{g}}{8r}e^{-m_{g}r}-\frac{M_{2}^{2}\left(1+c^{2}\right)^{2}m_{2}^{2}}{32c^{3}m_{g}^{4}r^{4}}e^{-2m_{g}r}\\ \nonumber
 &  & -\frac{\left(1+c^{2}\right)^{2}M_{2}^{2}m_{2}^{2}}{16c^{3}m_{g}^{3}r^{3}}e^{-2m_{g}r}+\frac{3M_{2}^{2}\left(1+c^{2}\right)\left[\left(5+4c^{2}\right)m_{1}^{2}-c^{2}m_{2}^{2}\right]}{64c^{3}m_{g}^{2}r^{2}}e^{-2m_{g}r}\\ \nonumber
 &  & +\frac{M_{2}^{2}\left(1+c^{2}\right)\left[\left(-25+44c^{2}\right)m_{1}^{2}+\left(-44+25c^{2}\right)m_{2}^{2}\right]}{128c^{3}m_{g}r}e^{m_{g}r}\mbox{Ei}\left(-3m_{g}r\right)+\frac{M_{1}M_{2}m_{g}}{2r}e^{m_{g}r}\mbox{Ei}\left(-2m_{g}r\right)\\ \nonumber
 &  & +\frac{M_{2}^{2}\left(1+c^{2}\right)\left[\left(25-44c^{2}\right)m_{1}^{2}+\left(44-25c^{2}\right)m_{2}^{2}\right]}{128c^{3}m_{g}r}e^{-m_{g}r}\mbox{Ei}\left(-m_{g}r\right)+\frac{3M_{1}M_{2}m_{g}^{2}}{8}\mbox{Ei}\left(-m_{g}r\right)\\ 
 &  & -\frac{M_{1}M_{2}m_{g}}{2r}e^{-m_{g}r}\log\left(\frac{r}{r_{0}}\right),
\end{eqnarray}

\begin{eqnarray}
\nonumber
\frac{\delta A_{2}}{G^{2}} & = & \frac{M_{1}^{2}}{2r^{2}}-\frac{M_{1}M_{2}}{c^{2}r^{2}}e^{-m_{g}r}-\frac{\left(1+c^{2}\right)M_{1}M_{2}}{2c^{2}m_{g}r^{3}}e^{-m_{g}r}-\frac{\left(1+c^{2}\right)M_{1}M_{2}}{2c^{2}m_{g}^{2}r^{4}}e^{-m_{g}r}\\ \nonumber
 &  & +\frac{M_{2}^{2}\left(1+c^{2}\right)^{2}\left(7m_{1}^{2}+8m_{2}^{2}\right)}{16c^{5}m_{g}^{4}r^{4}}e^{-2m_{g}r}+\frac{M_{2}^{2}\left(1+c^{2}\right)^{2}\left(7m_{1}^{2}+8m_{2}^{2}\right)}{8c^{5}m_{g}^{3}r^{3}}e^{-2m_{g}r}\\ \nonumber
 &  & +\frac{M_{2}^{2}\left(1+c^{2}\right)\left[\left(19+4c^{2}\right)m_{1}^{2}+\left(28+13c^{2}\right)m_{2}^{2}\right]}{32c^{5}m_{g}^{2}r^{2}}e^{-2m_{g}r}\\ \nonumber
 &  & -\frac{3M_{2}^{2}m_{g}}{8c^{2}r}e^{-2m_{g}r}+\frac{M_{2}^{2}\left(1+c^{2}\right)\left[\left(-25+44c^{2}\right)m_{1}^{2}+\left(-44+25c^{2}\right)m_{2}^{2}\right]}{64c^{5}m_{g}r}e^{m_{g}r}\mbox{Ei}\left(-3m_{g}r\right)\\ \nonumber
 &  & +\frac{M_{1}M_{2}m_{g}}{c^{2}r}e^{m_{g}r}\mbox{Ei}\left(-2m_{g}r\right)-\frac{3M_{2}^{2}m_{g}^{2}}{4c^{2}}\mbox{Ei}\left(-2m_{g}r\right)-\frac{M_{1}M_{2}m_{g}}{c^{2}r}e^{-m_{g}r}\log\left(\frac{r}{r_{0}}\right)\\
 &  & +\frac{M_{2}^{2}\left(1+c^{2}\right)\left[\left(25-44c^{2}\right)m_{1}^{2}+\left(44-25c^{2}\right)m_{2}^{2}\right]}{64c^{5}m_{g}r}e^{-m_{g}r}\mbox{Ei}\left(-m_{g}r\right),
\end{eqnarray}

\begin{eqnarray}
\nonumber
\frac{\delta B_{2}}{G^{2}} & = & \frac{M_{1}^{2}}{4r^{2}}+\frac{3M_{1}M_{2}m_{g}}{8r}e^{-m_{g}r}+\frac{M_{1}M_{2}}{8r^{2}}e^{-m_{g}r}-\frac{3\left(1+c^{2}\right)M_{1}M_{2}}{c^{2}m_{g}r^{3}}e^{-m_{g}r}-\frac{3\left(1+c^{2}\right)M_{1}M_{2}}{2c^{2}m_{g}^{2}r^{4}}e^{-m_{g}r}\\ \nonumber
 &  & +\frac{M_{2}^{2}\left(1+c^{2}\right)\left[-\left(32+97c^{2}+68c^{4}\right)m_{1}^{2}+\left(80+112c^{2}+29c^{4}\right)m_{2}^{2}\right]}{64c^{5}m_{g}^{2}r^{2}}e^{-2m_{g}r}\\ \nonumber
 &  & +\frac{M_{2}^{2}\left(1+c^{2}\right)^{2}\left[-\left(13+44c^{2}\right)m_{1}^{2}+4\left(15+7c^{2}\right)m_{2}^{2}\right]}{16c^{5}m_{g}^{3}r^{3}}e^{-2m_{g}r}\\ \nonumber
 &  & +\frac{M_{2}^{2}\left(1+c^{2}\right)^{2}\left[-2\left(17+24c^{2}\right)m_{1}^{2}+5\left(32+29c^{2}\right)m_{2}^{2}\right]}{32c^{5}m_{g}^{4}r^{4}}e^{-2m_{g}r}\\ \nonumber
 &  & +\frac{5M_{2}^{2}\left(1+c^{2}\right)^{3}m_{2}^{2}}{c^{5}m_{g}^{5}r^{5}}e^{-2m_{g}r}+\frac{5M_{2}^{2}\left(1+c^{2}\right)^{3}m_{2}^{2}}{2c^{5}m_{g}^{6}r^{6}}e^{-2m_{g}r}\\ \nonumber
 &  & +\frac{\left(1+c^{2}\right)^{2}M_{2}^{2}\left[\left(-25+44c^{2}\right)m_{1}^{2}+\left(-44+25c^{2}\right)m_{2}^{2}\right]}{128c^{5}m_{g}^{3}r^{3}}\times\\ \nonumber
& & \times\left[2-2m_{g}r+c^{2}\left(1+c^{2}\right)^{-1}m_{g}^{2}r^{2}\right]e^{m_{g}r}\mbox{Ei}\left(-3m_{g}r\right)\\ \nonumber
 &  & +\frac{\left(1+c^{2}\right)M_{1}M_{2}\left[2-2m_{g}r+c^{2}\left(1+c^{2}\right)^{-1}m_{g}^{2}r^{2}\right]}{2c^{2}m_{g}r^{3}}e^{m_{g}r}\mbox{Ei}\left(-2m_{g}r\right)+\frac{3M_{2}M_{1}m_{g}^{2}}{8}\mbox{Ei}\left(-m_{g}r\right)\\ \nonumber
 &  & -\frac{M_{2}^{2}\left(1+c^{2}\right)^{2}\left[\left(-25+44c^{2}\right)m_{1}^{2}+\left(-44+25c^{2}\right)m_{2}^{2}\right]}{128c^{5}m_{g}^{3}r^{3}}\\ \nonumber
& & \times\left[2+2m_{g}r+c^{2}\left(1+c^{2}\right)^{-1}m_{g}^{2}r^{2}\right]e^{-m_{g}r}\mbox{Ei}\left(-m_{g}r\right)\\
 &  & -\frac{\left(1+c^{2}\right)M_{1}M_{2}\left[2+2m_{g}r+c^{2}\left(1+c^{2}\right)^{-1}m_{g}^{2}r^{2}\right]}{2c^{2}m_{g}r^{3}}e^{-m_{g}r}\log\left(\frac{r}{r_{0}}\right),
\end{eqnarray}

\begin{eqnarray}
\nonumber
\frac{\delta H_{2}}{G^{2}} & = & \frac{M_{1}^{2}}{4r^{2}}+\frac{3M_{1}M_{2}m_{g}}{8r}e^{-m_{g}r}-\frac{\left(4+3c^{2}\right)M_{1}M_{2}}{8c^{2}r^{2}}e^{-m_{g}r}+\frac{7\left(1+c^{2}\right)M_{1}M_{2}}{4c^{2}m_{g}r^{3}}e^{-m_{g}r}\\ \nonumber
 &  & +\frac{M_{1}M_{2}\left(1+c^{2}\right)}{c^{2}m_{g}^{2}r^{4}}e^{-m_{g}r}-\frac{M_{2}^{2}\left(1+c^{2}\right)\left[\left(19+16c^{2}\right)m_{1}^{2}+\left(4+c^{2}\right)m_{2}^{2}\right]}{64c^{5}m_{g}^{2}r^{2}}e^{-2m_{g}r}\\ \nonumber
 &  & +\frac{M_{2}^{2}\left(1+c^{2}\right)^{2}\left[\left(-17+28c^{2}\right)m_{1}^{2}-\left(44+c^{2}\right)m_{2}^{2}\right]}{32c^{5}m_{g}^{3}r^{3}}e^{-2m_{g}r}-\frac{M_{2}^{2}\left(1+c^{2}\right)^{3}m_{2}^{2}}{2c^{5}m_{g}^{6}r^{6}}e^{-2m_{g}r}\\ \nonumber
 &  & +\frac{M_{2}^{2}\left(1+c^{2}\right)^{2}\left[2\left(1+8c^{2}\right)m_{1}^{2}-\left(40+27c^{2}\right)m_{2}^{2}\right]}{32c^{5}m_{g}^{4}r^{4}}e^{-2m_{g}r}-\frac{M_{2}^{2}\left(1+c^{2}\right)^{3}m_{2}^{2}}{c^{5}m_{g}^{5}r^{5}}e^{-2m_{g}r}\\ \nonumber
 &  & +\frac{\left(1+c^{2}\right)^{2}M_{2}^{2}\left[\left(-25+44c^{2}\right)m_{1}^{2}+\left(-44+25c^{2}\right)m_{2}^{2}\right]}{128c^{6}m_{g}^{3}r^{3}}\times\\ \nonumber
 & & \times\left[-c\left(1+c^{2}\right)^{-1}m_{g}^{2}r^{2}+c\left(-1+m_{g}r\right)\right]e^{m_{g}r}\mbox{Ei}\left(-3m_{g}r\right)\\ \nonumber 
 &  & +\frac{\left(1+c^{2}\right)M_{1}M_{2}\left[-c\left(1+c^{2}\right)^{-1}m_{g}^{2}r^{2}+c\left(-1+m_{g}r\right)\right]}{2c^{3}m_{g}r^{3}}e^{m_{g}r}\mbox{Ei}\left(-2m_{g}r\right)\\ \nonumber
 &  & +\frac{\left(1+c^{2}\right)^{2}M_{2}^{2}\left(44c^{2}m_{1}^{2}-25m_{1}^{2}+25c^{2}m_{2}^{2}-44m_{2}^{2}\right)}{128c^{6}m_{g}^{3}r^{3}}\times\\ \nonumber
 & & \times\left[c\left(1+c^{2}\right)^{-1}m_{g}^{2}r^{2}+c\left(1+m_{g}r\right)\right]e^{-m_{g}r}\mbox{Ei}\left(-m_{g}r\right)+\frac{3M_{2}M_{1}m_{g}^{2}}{8}\mbox{Ei}\left(-m_{g}r\right)\\
 &  & +\frac{\left(1+c^{2}\right)M_{1}M_{2}\left[c\left(1+m_{g}r\right)+c\left(1+c^{2}\right)^{-1}m_{g}^{2}r^{2}\right]}{2c^{3}m_{g}r^{3}}e^{-m_{g}r}\log\left(\frac{r}{r_{0}}\right).
\end{eqnarray}

Here
\begin{equation}
m_{1}^{2}\equiv m^{2}\left(\beta_{1}+\beta_{2}c\right),\quad m_{2}^{2}\equiv m^{2}\left(\beta_{2}c+\beta_{3}c^{2}\right),
\end{equation}
\begin{equation}
m_{g}^{2}=\left(c^{-1}+c\right)\left(m_{1}^{2}+m_{2}^{2}\right)
\end{equation}
and
\begin{equation}
\mbox{Ei}\left(x\right)=-\intop_{-x}^{\infty}\frac{e^{-t}}{t}dt.
 \end{equation}

\section{Velocity dispersions\label{sec:Velocity-dispersions}}

Since we can decompose the gravitational potentials for massive and massless particles as $\Phi=\Phi_{GR}+\Phi_{Y}$ and $\varphi=\varphi_{GR}+\varphi_{Y}$, where the subscript $GR$ denotes the general relativity terms and $Y$ the Yukawa terms of the potentials, and
both the velocity dispersion and gravitational lensing angle depends linearly on these potentials, we can decompose also these as
$\sigma_r=\sigma_{GR,r}+\sigma_{Y,r}$ and $\hat\alpha=\hat\alpha_{GR}+\hat\alpha_{Y}$. 

The radially dependent velocity dispersion is given by eq.~(\ref{eq:sigmar2}),
\begin{equation}
\sigma_{r}^{2}(r)=\frac{1}{\nu r^{2\zeta}}\intop_{r}^{\infty}\nu r^{2\zeta}\Phi^{\prime}dr.
\end{equation}
The observed velocity dispersion, $\sigma_{\star}^{2}$, is then given by a line-of-sight luminosity weighted average over the spectroscopic aperture of size $R_{max}$
\begin{equation}\label{eq:sigstar}
\sigma_{\star}^{2}=\frac{\intop_{0}^{R_{max}}dRRw(R)\intop_{-\infty}^{\infty}dZ\nu(r)(1-\zeta\frac{R^{2}}{r^{2}})\sigma_{r}^{2}}{\intop_{0}^{R_{max}}dRRw(R)\intop_{-\infty}^{\infty}dZ\nu(r)},
\end{equation}
where $Z^{2}=r^{2}-R^{2}$ and
\begin{equation}
w(R)=e^{-R^{2}/2\bar{\sigma}_{atm}^{2}}
\end{equation}
is the aperture weighting function.

\subsection{General relativity term}

For the GR term in the velocity dispersion, we can substitute $\Phi^{\prime}$ with $GM(r)/r^{2}$ through use of Poisson's equation, giving 
\begin{equation}
\sigma_{GR,r}^{2}=\frac{G}{\nu r^{2\zeta}}\intop_{r}^{\infty}\nu r^{2\zeta-2}M(r)dr=\frac{G}{r^{2\zeta-\delta}}\intop_{r}^{\infty}r^{2\zeta-\delta-2}M(r)dr,
\end{equation}
where 
\begin{equation}
M(r)=\intop_{0}^{r}4\pi r^{2}\rho(r)dr=4\pi\rho_{0}r_{0}^{\gamma}\intop_{0}^{r}r^{2-\gamma}dr.
\end{equation}
For $\gamma<3$, we obtain 
\begin{equation}
M(r)=\frac{4\pi\rho_{0}r_{0}^{\gamma}}{(3-\gamma)}r^{3-\gamma}
\end{equation}
and 
\begin{equation}
\sigma_{GR,r}^{2}=\frac{Gr_{0}^{\gamma}}{(3-\gamma)r^{2\zeta-\delta}}\intop_{r}^{\infty}r^{2\zeta-\delta-\gamma+1}dr=\frac{4\pi G\rho_{0}r_{0}^{\gamma}}{(3-\gamma)(\gamma+\delta-2\zeta-2)}r^{2-\gamma},
\end{equation}
for $2\zeta-\delta-\gamma<-2$. The Singular Isothermal Sphere model (SIS) is given by $\zeta=0$ and $\gamma=\delta=2$, giving $\sigma_{GR,r}^{2}=2\pi G\rho_{0}r_{0}^{2}$. 
The GR term in the observed velocity dispersion is now given by (obtained by changing variables of the inner integrals of eq.~(\ref{eq:sigstar}) to $x=R/r$)
\begin{eqnarray}
\sigma_{GR,\star}^{2} & = & \frac{4\pi G\rho_{0}r_{0}^{\gamma}}{(3-\gamma)(\gamma+\delta-2\zeta-2)}\frac{\left[\lambda(\gamma+\delta-2)-\zeta\lambda(\gamma+\delta)\right]}{\lambda(\delta)}\times\nonumber \\
 &  & \frac{\intop_{0}^{R_{max}}dRR^{4-\gamma-\delta}w(R)}{\intop_{0}^{R_{max}}dRR^{2-\delta}w(R)}.
\end{eqnarray}
Here, the lambda-function $\lambda(x)=\Gamma[(x-1)/2]/\Gamma(x/2)$ where $\Gamma$ is the gamma-function. 
This can be solved analytically if put $R_{max}=\infty$ to get
\begin{equation}
\sigma_{GR,\star}^{2}=\frac{4\pi G\rho_{0}r_{0}^{\gamma}(2\bar{\sigma}_{atm}^{2})^{1-\gamma/2}}{(3-\gamma)(\gamma+\delta-2\zeta-2)}\frac{\left[\lambda(\gamma+\delta-2)-\zeta\lambda(\gamma+\delta)\right]}{\lambda(\delta)}\frac{\Gamma\left(\frac{5-\gamma-\delta}{2}\right)}{\Gamma\left(\frac{3-\delta}{2}\right)}.
\end{equation}
We note that for $\zeta=0$ and $\gamma=\delta=2$, we get back $\sigma_{GR,\star}^{2}=\sigma_{GR,r}^{2}=2\pi G\rho_{0}r_{0}^{2}$, as expected. 

\subsection{Yukawa term}

Now, in principle we can derive the velocity dispersion $\sigma_{Y,r}^{2}(r)$ and $\sigma_{Y,\star}^{2}$ corresponding to the Yukawa term in the potential.
First, we need the Yukawa term in the gravitational potential for the case of a spherically symmetric mass distribution. For a potential of the form 
\begin{equation}
\Phi_{Y}=-\frac{4c^2M}{3r}e^{-m_{g}r},
\end{equation}
if the mass $M=4\pi R^{2}\rho(R)dR$ is distributed in a thin shell of radius $R$, the corresponding potential is 
\begin{equation}
\Phi_{Y}(r,R)=-\frac{k4\pi R\rho(R)dR}{m_{g}r}\times\left\{ \begin{aligned}e^{-m_{g}r}\sinh(m_{g}R)\quad r\geq R\\
e^{-m_{g}R}\sinh(m_{g}r)\quad r\leq R
\end{aligned}
.\right.
\end{equation}
In order to get the total  potential from a spherically symmetric matter distribution, we integrate over a series of shells
\begin{eqnarray}
 &\Phi_{Y}(r)=\intop_{0}^{\infty}\Phi_{Y}(r,R)dR=-\frac{k4\pi}{m_{g}r}\times\nonumber \\
 & \left\{ e^{-m_{g}r}\intop_{o}^{r}\sinh(m_{g}R)R\rho(R)dR+\sinh(m_{g}r)\intop_{r}^{\infty}e^{-m_{g}R}R\rho(R)dR\right\} \,.\label{eq:Phi1}
\end{eqnarray}
Next, we differentiate with respect to $r$, 
\begin{equation}
\Phi_{Y}^{\prime}=\intop_{0}^{\infty}\Phi_{Y}^{\prime}dR=m_{g}\left[-\Phi_{Y}(1+\frac{1}{m_{g}r})-\frac{k4\pi}{m_{g}r}e^{m_{g}r}\intop_{r}^{\infty}e^{-m_{g}R}R\rho(R)dR\right].
\end{equation}
In the simplest case of $\zeta=0$ and $\gamma=\delta=2$, and $k=(4c^{2})/3$ we get
\begin{equation}
\Phi_{Y}(r)=-\frac{4c^{2}\Phi_{GR}^{\prime}}{3x}\left\{ e^{-x}{\rm Shi}(x)-\sinh(x){\rm Ei}(-x)\right\} ,
\end{equation}
and 
\begin{equation}
\Phi_{Y}^{\prime}=\frac{4c^{2}\Phi_{GR}^{\prime}}{3x}\left\{ (1+x)\left[e^{-x}{\rm Shi}(x)-\sinh(x){\rm Ei}(-x)\right]+xe^{x}Ei(-x)\right\} ,
\end{equation}
where $x\equiv m_{g}r$, $\Phi_{GR}^{\prime}=4\pi\rho_{0}r_{0}^{2}/r$, ${\rm Shi(x)}$ is the hyperbolic sine integral function and ${\rm Ei(x)}$ is the exponential integral function. 

Since the derived expressions do not render the observed Yukawa part of the velocity dispersion, $\sigma_{Y,\star}^{2}$, analytically solvable, we use the following approximation: Since the observed velocity dispersion is a weighted average over a few spectroscopic apertures (the only scale in the problem since the luminosity and matter profiles are given by pure power laws), we can employ a constant correction to $\sigma_{r}^{2}$ given by the correction to $\Phi$ at a distance equal to $r=r_{s}=\bar{\sigma}_{atm}$, i.e.
\begin{equation}
\sigma_{r}^{2}=\sigma_{GR,r}^{2}\left[1+\frac{4c^{2}}{3}e^{-m_{g}r_{s}}\right],
\end{equation}
and 
\begin{equation}
\sigma_{\star}^{2}=\sigma_{GR,\star}^{2}\left[1+\frac{4c^{2}}{3}e^{-m_{g}r_{s}}\right].
\end{equation}
It can be shown numerically that this approximation gives a maximum fractional error of the derived velocity dispersion of $\sim 12\,\%$ from the exact value, assuming $c=1$.
 
\section{Gravitational lensing\label{sec:Gravitational-lensing}}

The gravitational deflection angle is given by eq.~(\ref{eq:alpha})
\begin{equation}
\hat{\alpha}=2\intop_{-\infty}^{\infty}\nabla_{\bot}\varphi dl.
\end{equation}
We make use of a scaled deflection angle $\alpha\equiv D_{ls}/D_{s}\hat{\alpha}$, where $D_{ls}$ and $D_{s}$ are angular diameter distances between the lens and the source and the observer and source, respectively. The scaled deflection angle fulfills the (spherically symmetric) lens equation 
\begin{equation}
\beta=\theta-\alpha,
\end{equation}
where $\theta$ is the angular position of the image with respect to the center of the deflector and $\beta$ is the angular position the source would have in absence of the lens (not to be confused with the $\beta_i$:s of the Lagrangian defining the HR theory). The scaled deflection angle can now be computed as 
\begin{equation}
\alpha(\theta)=\frac{1}{\pi}\int_{0}^{\infty}\kappa(x)xdx\int_{0}^{2\pi}\frac{(\theta-x\cos\eta)d\eta}{\theta^{2}+x^{2}-2\theta x\cos\eta},
\end{equation}
where $\kappa(\theta)$ is the scaled surface mass density
\begin{equation}
\kappa(\theta)=\frac{\Sigma(\theta D_{l})}{\Sigma_{cr}},
\end{equation}
where 
\begin{equation}
\Sigma(R)=\intop_{-\infty}^{\infty}\rho(\sqrt{R^{2}+l^{2}})dl
\end{equation}
and 
\begin{equation}
\Sigma_{cr}=\frac{1}{4\pi G}\frac{D_{s}}{D_{l}D_{ls}}.
\end{equation}
Here, $D_{l}$ is the angular diameter distance to the lens. 

\subsection{General relativity term}

In GR, it can be shown that the deflection angle is given by
\begin{equation}
\alpha_{GR}=\frac{4Gm(R)}{c^{2}R}
\end{equation}
where $m(R)$ is the projected mass enclosed within radius $R$. For the power law density profile, we begin by computing the surface mass density
\begin{equation}
\Sigma(R)=\intop_{-\infty}^{\infty}\rho(\sqrt{R^{2}+l^{2}})dl=2\rho_{0}\intop_{0}^{\infty}(R^{2}+l^{2})^{-\gamma/2}dl=\sqrt{\pi}\lambda(\gamma)\rho_{0}R^{1-\gamma}.
\end{equation}
Now 
\begin{equation}
m(R)=\intop_{0}^{R}\Sigma(R)dA=2\pi^{3/2}\lambda(\gamma)\rho_{0}r_{0}^{\gamma}\intop R^{2-\gamma}dR=\frac{2\pi^{3/2}\lambda(\gamma)\rho_{0}r_{0}^{\gamma}}{3-\gamma}R^{3-\gamma}
\end{equation}
and
\begin{equation}
\alpha_{GR}(R)=\frac{8G\pi^{3/2}\lambda(\gamma)\rho_{0}r_{0}^{\gamma}}{(3-\gamma)}R^{2-\gamma}.
\end{equation}
For $\gamma=2$, we get
\begin{equation}
\alpha=8G\pi^{2}\rho_{0}r_{0}^{2}=4\pi\sigma_{GR,r}^{2}.
\end{equation}

\subsection{Yukawa term}

Using eq.~(\ref{eq:alpha}), we can show that the deflection from the Yukawa term in the potential in units of the deflection angle from the GR term is given by
\begin{equation}
\frac{\alpha_{Y}}{\alpha_{GR}}=c^{2}B^{2}\intop_{0}^{\infty}\frac{\exp(-\sqrt{B^{2}+x^{2}})}{(B^{2}+x^{2})^{3/2}}\left(\sqrt{B^{2}+x^{2}}+1\right)dx,
\end{equation}
where \textbf{$B\equiv m_{g}b$} is the impact parameter in units of $m_{g}^{-1}$. This is not analytically solvable, but a fit to this function gives
\begin{equation}
\frac{\alpha_{Y}}{c^{2}\alpha_{GR}}=\frac{q}{q+B^{2}},\label{eq:ydfit}
\end{equation}
where $q\simeq1.45$. 
Writing the total scaled deflection angle as $\alpha=\alpha_{GR}+\alpha_{Y}$, we can write
\begin{eqnarray}
\alpha_{Y}(\theta) & = & \frac{c^{2}}{\pi}\int_{0}^{\infty}\kappa(x)xdx\,\times\\
 &  & \int_{0}^{2\pi}\frac{(\theta-x\cos\eta)}{\theta^{2}+x^{2}-2\theta x\cos\eta}\frac{d\eta}{1+\frac{D_{l}^{2}m_{g}^{2}}{q}(\theta^{2}+x^{2}-2\theta x\cos\eta)}.
\end{eqnarray}
The inner integral over angle $\eta$ can be shown to equal
\begin{equation}
\frac{\pi}{\theta}\times\left\{ \begin{aligned}g(z,z^{\prime})+1\quad z\geq z^{\prime}\\
g(z,z^{\prime})-1\quad z\leq z^{\prime}
\end{aligned}
,\right.
\end{equation}
where
\begin{equation}
g(z,z^{\prime})=\frac{1-z^{2}+z^{\prime2}}{\sqrt{z^{4}-2z^{2}(z^{\prime2}-1)+(z^{\prime2}+1)^{2}}},
\end{equation}
and 
\begin{equation}
z\equiv\frac{m_{g}D_{l}\theta}{\sqrt{q}},\quad z^{\prime}\equiv\frac{m_{g}D_{l}x}{\sqrt{q}}.
\end{equation}
Given the power law density profile, we can show that
\begin{equation}
\alpha=\alpha_{GR}\left(1+\frac{\alpha_{Y}}{\alpha_{GR}}\right)=\alpha_{GR}\left[1+\frac{c^{2}(3-\gamma)}{2z^{3-\gamma}}h(z,\gamma)\right]
\end{equation}
where
\begin{equation}
h(z,\gamma)=\int_{0}^{z}(g+1)z^{\prime2-\gamma}dz^{\prime}+\int_{z}^{\infty}(g-1)z^{\prime2-\gamma}dz^{\prime},
\end{equation}
and
\begin{equation}
\alpha_{GR}=\frac{8G\pi^{3/2}\lambda(\gamma)\rho_{0}r_{0}^{\gamma}}{(3-\gamma)}\theta^{2-\gamma}.
\end{equation}
To simplify the analysis, we again assume that the correction can be approximated by a constant rescaling of the lensing potential of $\varphi_{Y}=\varphi_{GR}e^{-z_{E}}$, giving 
\begin{equation}
\alpha=\alpha_{GR}\left(1+\frac{\alpha_{Y}}{\alpha_{GR}}\right)=\alpha_{GR}\left[1+c^{2}e^{-z_{E}}\right],
\end{equation}
where $z_{E}=m_{g}D_{l}\theta_{E}/\sqrt{q}$ and $\theta_E$ is the Einstein radius of the system. 
Numerical calculations show that this gives a fractional error of the deflection angle of at most $6\,\%$, for $c=1$.
Since $\theta\simeq\theta_E$, we can write the lens equation $\theta_E=\alpha(\rho_{0}r_{0}^{\gamma},\gamma,\theta)$ in terms of $\theta_E$ 
\begin{equation}
\theta_{E}=\frac{8G\pi^{3/2}\lambda(\alpha)\rho_{0}r_{0}^{\gamma}}{(3-\gamma)}\theta_{E}^{2-\gamma}\left[1+c^{2}e^{-z_{E}}\right]\label{eq:theta_E}.
\end{equation}

\section{Fitting to the data}

From eq.~(\ref{eq:theta_E}), given a measured $\theta_{E}$ and assuming values for $\gamma$, $m_{g}$ and $c$, we can solve for $\rho_{0}r_{0}^{\gamma}$. This is then put into the expression for the observed velocity dispersion $\sigma_{\star}^{2}$ to give 
\begin{equation}
\sigma_{\star}^{2}=\sigma_{GR,\star}^{2}\frac{1+\frac{4c^{2}}{3}e^{-m_{g}r_{s}}}{1+c^{2}e^{-z_{E}}},
\end{equation}
where
\begin{eqnarray}
\sigma_{GR,\star}^{2} & = & \frac{\theta_{E}^{\gamma-1}}{2\sqrt{\pi}(\gamma+\delta-2\zeta-2)}\frac{\left[\lambda(\gamma+\delta-2)-\zeta\lambda(\gamma+\delta)\right]}{\lambda(\gamma)\lambda(\delta)}\times\nonumber \\
 &  & \frac{\intop_{0}^{R_{max}}dRR^{4-\gamma-\delta}w(R)}{\intop_{0}^{R_{max}}dRR^{2-\delta}w(R)}.
\end{eqnarray}
The computed value of $\sigma_{\star}$ can then be compared to the observed value in order to constrain the parameters of the model. Now, since the approximations employed when calculating the velocity dispersion and lensing deflection angle are correlated, it can be shown that when combined, the maximal total fractional error on the derived velocity dispersion when normalized using the lensing deflection angle, is always less than 10\,\%. (This error is largest when $\lambda_g\sim r_g$ and goes to zero as $\lambda_g\rightarrow 0$ or $\lambda_g\rightarrow\infty$). Although this error is comparable to the observational errors, it will have a negligible effect on the derived constraints on $\lambda_g$ and $c$. 

\clearpage

\bibliographystyle{JHEP}
\bibliography{slcbmg}{}

\providecommand{\href}[2]{#2}\begingroup\raggedright\begin{thebibliography}{10}

\bibitem{HassanRosen2012b}
S.~F. {Hassan} and R.~A. {Rosen}, {\it {Bimetric gravity from ghost-free
  massive gravity}},  {\em Journal of High Energy Physics} {\bf 2} (2012) 126,
  [\href{http://xxx.lanl.gov/abs/1109.3515}{{\tt arXiv:1109.3515}}].

\bibitem{HassanRosen2012c}
S.~F. {Hassan} and R.~A. {Rosen}, {\it {Confirmation of the secondary
  constraint and absence of ghost in massive gravity and bimetric gravity}},
  {\em Journal of High Energy Physics} {\bf 4} (2012) 123,
  [\href{http://xxx.lanl.gov/abs/1111.2070}{{\tt arXiv:1111.2070}}].

\bibitem{DeFelice2013}
A.~De~Felice, T.~Nakamura, and T.~Tanaka, {\it {Possible existence of viable
  models of bi-gravity with detectable graviton oscillations by gravitational
  wave detectors}},  \href{http://xxx.lanl.gov/abs/1304.3920}{{\tt
  arXiv:1304.3920}}.

\bibitem{Volkov:2013roa}
M.~S. Volkov, {\it {Self-accelerating cosmologies and hairy black holes in
  ghost-free bigravity and massive gravity}},
  \href{http://xxx.lanl.gov/abs/1304.0238}{{\tt arXiv:1304.0238}}.

\bibitem{vonStrauss:2011mq}
M.~von Strauss, A.~Schmidt-May, J.~Enander, E.~Mortsell, and S.~Hassan, {\it
  {Cosmological Solutions in Bimetric Gravity and their Observational Tests}},
  {\em Journal of Cosmology and Astroparticle Physics} {\bf 1203} (2012) 042,
  [\href{http://xxx.lanl.gov/abs/1111.1655}{{\tt arXiv:1111.1655}}].

\bibitem{Maeda:2013bha}
K.~Maeda and M.~S. Volkov, {\it {Anisotropic universes in the ghost-free
  bigravity}},  \href{http://xxx.lanl.gov/abs/1302.6198}{{\tt
  arXiv:1302.6198}}.

\bibitem{Volkov:2011an}
M.~S. Volkov, {\it {Cosmological solutions with massive gravitons in the
  bigravity theory}},  {\em Journal of High Energy Physics} {\bf 1201} (2012)
  035, [\href{http://xxx.lanl.gov/abs/1110.6153}{{\tt arXiv:1110.6153}}].

\bibitem{Volkov:2012zb}
M.~S. Volkov, {\it {Exact self-accelerating cosmologies in the ghost-free
  massive gravity -- the detailed derivation}},  {\em Physical Review} {\bf
  D86} (2012) 104022, [\href{http://xxx.lanl.gov/abs/1207.3723}{{\tt
  arXiv:1207.3723}}].

\bibitem{Capozziello:2012re}
S.~Capozziello and P.~Martin-Moruno, {\it {Bounces, turnarounds and
  singularities in bimetric gravity}},  {\em Physics Letters} {\bf B719} (2013)
  14--17, [\href{http://xxx.lanl.gov/abs/1211.0214}{{\tt arXiv:1211.0214}}].

\bibitem{Akrami:2012vf}
Y.~Akrami, T.~S. Koivisto, and M.~Sandstad, {\it {Accelerated expansion from
  ghost-free bigravity: a statistical analysis with improved generality}},
  {\em Journal of High Energy Physics} {\bf 1303} (2013) 099,
  [\href{http://xxx.lanl.gov/abs/1209.0457}{{\tt arXiv:1209.0457}}].

\bibitem{Akrami:2013ffa}
Y.~Akrami, T.~S. Koivisto, D.~F. Mota, and M.~Sandstad, {\it {Bimetric gravity
  doubly coupled to matter: theory and cosmological implications}},
  \href{http://xxx.lanl.gov/abs/1306.0004}{{\tt arXiv:1306.0004}}.

\bibitem{Berg:2012kn}
M.~Berg, I.~Buchberger, J.~Enander, E.~Mortsell, and S.~Sjors, {\it {Growth
  Histories in Bimetric Massive Gravity}},  {\em Journal of Cosmology and
  Astroparticle Physics} {\bf 1212} (2012) 021,
  [\href{http://xxx.lanl.gov/abs/1206.3496}{{\tt arXiv:1206.3496}}].

\bibitem{Comelli:2012db}
D.~Comelli, M.~Crisostomi, and L.~Pilo, {\it {Perturbations in Massive Gravity
  Cosmology}},  {\em Journal of High Energy Physics} {\bf 1206} (2012) 085,
  [\href{http://xxx.lanl.gov/abs/1202.1986}{{\tt arXiv:1202.1986}}].

\bibitem{Sakakihara:2012iq}
Y.~Sakakihara, J.~Soda, and T.~Takahashi, {\it {On Cosmic No-hair in Bimetric
  Gravity and the Higuchi Bound}},  {\em PTEP} {\bf 2013} (2013) 033E02,
  [\href{http://xxx.lanl.gov/abs/1211.5976}{{\tt arXiv:1211.5976}}].

\bibitem{Khosravi:2012rk}
N.~Khosravi, H.~R. Sepangi, and S.~Shahidi, {\it {Massive cosmological scalar
  perturbations}},  {\em Physical Review} {\bf D86} (2012) 043517,
  [\href{http://xxx.lanl.gov/abs/1202.2767}{{\tt arXiv:1202.2767}}].

\bibitem{Kuhnel:2012gh}
F.~Kuhnel, {\it {On Instability of Certain Bi-Metric and Massive-Gravity
  Theories}},  \href{http://xxx.lanl.gov/abs/1208.1764}{{\tt arXiv:1208.1764}}.

\bibitem{deRhamGabadadze2010}
C.~de~Rham and G.~Gabadadze, {\it {Generalization of the Fierz-Pauli Action}},
  {\em Physical Review} {\bf D82} (2010) 044020,
  [\href{http://xxx.lanl.gov/abs/1007.0443}{{\tt arXiv:1007.0443}}].

\bibitem{dRGT2010}
C.~de~Rham, G.~Gabadadze, and A.~J. Tolley, {\it {Resummation of Massive
  Gravity}},  {\em Physical Review Letters} {\bf 106} (2011) 231101,
  [\href{http://xxx.lanl.gov/abs/1011.1232}{{\tt arXiv:1011.1232}}].

\bibitem{HassanRosen2011}
S.~F. {Hassan} and R.~A. {Rosen}, {\it {On non-linear actions for massive
  gravity}},  {\em Journal of High Energy Physics} {\bf 7} (2011) 9,
  [\href{http://xxx.lanl.gov/abs/1103.6055}{{\tt arXiv:1103.6055}}].

\bibitem{HassanRosen2012a}
S.~F. {Hassan} and R.~A. {Rosen}, {\it {Resolving the Ghost Problem in
  Nonlinear Massive Gravity}},  {\em Physical Review Letters} {\bf 108} (2012),
  no.~4 041101, [\href{http://xxx.lanl.gov/abs/1106.3344}{{\tt
  arXiv:1106.3344}}].

\bibitem{HassanRosenSchmidtMay2012}
S.~F. {Hassan}, R.~A. {Rosen}, and A.~{Schmidt-May}, {\it {Ghost-free massive
  gravity with a general reference metric}},  {\em Journal of High Energy
  Physics} {\bf 2} (2012) 26, [\href{http://xxx.lanl.gov/abs/1109.3230}{{\tt
  arXiv:1109.3230}}].

\bibitem{Fierz1939}
M.~{Fierz}, {\it {\"Uber die relativistische Theorie kr\"afterfreier Teilchen
  mit beliebigem Spin}},  {\em Helvetica Physica Acta} {\bf 12} (1939) 3--37.

\bibitem{FierzPauli1939}
M.~{Fierz} and W.~{Pauli}, {\it {On Relativistic Wave Equations for Particles
  of Arbitrary Spin in an Electromagnetic Field}},  {\em Royal Society of
  London Proceedings Series A} {\bf 173} (1939) 211--232.

\bibitem{IshamSalamStrathdee1971}
C.~J. Isham, A.~Salam, and J.~A. Strathdee, {\it {F-dominance of gravity}},
  {\em Physical Review} {\bf D3} (1971) 867--873.

\bibitem{SalamStrathdee1976}
A.~Salam and J.~Strathdee, {\it {A Class of Solutions for the Strong Gravity
  Equations}},  {\em Physical Review} {\bf D16} (1977) 2668.

\bibitem{ChamseddineSalamStrathdee1978}
A.~H. Chamseddine, A.~Salam, and J.~Strathdee, {\it {Strong Gravity and
  Supersymmetry}},  {\em Nuclear Physics} {\bf B136} (1978) 248--258.

\bibitem{ArkaniHamed2003}
N.~{Arkani-Hamed}, H.~{Georgi}, and M.~D. {Schwartz}, {\it {Effective field
  theory for massive gravitons and gravity in theory space}},  {\em Annals of
  Physics} {\bf 305} (2003) 96--118,
  [\href{http://xxx.lanl.gov/abs/hep-th/02}{{\tt hep-th/02}}].

\bibitem{Creminelli2005}
P.~Creminelli, A.~Nicolis, M.~Papucci, and E.~Trincherini, {\it {Ghosts in
  massive gravity}},  {\em Journal of High Energy Physics} {\bf 0509} (2005)
  003, [\href{http://xxx.lanl.gov/abs/hep-th/0505147}{{\tt hep-th/0505147}}].

\bibitem{Groot2007}
S.~{Groot Nibbelink}, M.~{Peloso}, and M.~{Sexton}, {\it {Nonlinear properties
  of vielbein massive gravity}},  {\em European Physical Journal C} {\bf 51}
  (2007) 741--752, [\href{http://xxx.lanl.gov/abs/hep-th/06}{{\tt hep-th/06}}].

\bibitem{Comelli:2011wq}
D.~Comelli, M.~Crisostomi, F.~Nesti, and L.~Pilo, {\it {Spherically Symmetric
  Solutions in Ghost-Free Massive Gravity}},  {\em Physical Review} {\bf D85}
  (2012) 024044, [\href{http://xxx.lanl.gov/abs/1110.4967}{{\tt
  arXiv:1110.4967}}].

\bibitem{Volkov:2012wp}
M.~S. Volkov, {\it {Hairy black holes in the ghost-free bigravity theory}},
  {\em Physical Review} {\bf D85} (2012) 124043,
  [\href{http://xxx.lanl.gov/abs/1202.6682}{{\tt arXiv:1202.6682}}].

\bibitem{Babichev:2013una}
E.~Babichev and A.~Fabbri, {\it {Instability of black holes in massive
  gravity}},  \href{http://xxx.lanl.gov/abs/1304.5992}{{\tt arXiv:1304.5992}}.

\bibitem{Hassan:2012wr}
S.~Hassan, A.~Schmidt-May, and M.~von Strauss, {\it {On Consistent Theories of
  Massive Spin-2 Fields Coupled to Gravity}},  {\em JHEP} {\bf 1305} (2013)
  086, [\href{http://xxx.lanl.gov/abs/1208.1515}{{\tt arXiv:1208.1515}}].

\bibitem{rotcurve}
J.~{Enander} and E.~{Mortsell}, {\it {In preparation}}, .

\bibitem{Vainshtein1972}
A.~I. {Vainshtein}, {\it {To the problem of nonvanishing gravitation mass}},
  {\em Physics Letters B} {\bf 39} (1972) 393--394.

\bibitem{Babichev:2013usa}
E.~Babichev and C.~Deffayet, {\it {An introduction to the Vainshtein
  mechanism}},  \href{http://xxx.lanl.gov/abs/1304.7240}{{\tt
  arXiv:1304.7240}}.

\bibitem{Babichev:2013pfa}
E.~Babichev and M.~Crisostomi, {\it {Restoring General Relativity in massive
  bi-gravity theory}},  \href{http://xxx.lanl.gov/abs/1307.3640}{{\tt
  arXiv:1307.3640}}.

\bibitem{2010ApJ...708..750S}
J.~{Schwab}, A.~S. {Bolton}, and S.~A. {Rappaport}, {\it {Galaxy-Scale
  Strong-Lensing Tests of Gravity and Geometric Cosmology: Constraints and
  Systematic Limitations}},  {\em The Astrophysical Journal} {\bf 708} (2010)
  750--757, [\href{http://xxx.lanl.gov/abs/0907.4992}{{\tt arXiv:0907.4992}}].

\bibitem{2011arXiv1111.5961S}
S.~{Sj\"ors} and E.~{M\"ortsell}, {\it {Spherically Symmetric Solutions in
  Massive Gravity and Constraints from Galaxies}},
  \href{http://xxx.lanl.gov/abs/1111.5961}{{\tt arXiv:1111.5961}}.

\bibitem{2008ApJ...682..964B}
A.~S. {Bolton}, S.~{Burles}, L.~V.~E. {Koopmans}, T.~{Treu}, R.~{Gavazzi},
  L.~A. {Moustakas}, R.~{Wayth}, and D.~J. {Schlegel}, {\it {The Sloan Lens ACS
  Survey. V. The Full ACS Strong-Lens Sample}},  {\em The Astrophysical
  Journal} {\bf 682} (2008) 964--984,
  [\href{http://xxx.lanl.gov/abs/0805.1931}{{\tt arXiv:0805.1931}}].

\bibitem{2008A&A...477..397G}
C.~{Grillo}, M.~{Lombardi}, and G.~{Bertin}, {\it {Cosmological parameters from
  strong gravitational lensing and stellar dynamics in elliptical galaxies}},
  {\em Astronomy \& Astrophysics} {\bf 477} (2008) 397--406,
  [\href{http://xxx.lanl.gov/abs/0711.0882}{{\tt arXiv:0711.0882}}].

\bibitem{2011PhRvD..83j3516D}
C.~{de Rham}, G.~{Gabadadze}, L.~{Heisenberg}, and D.~{Pirtskhalava}, {\it
  {Cosmic acceleration and the helicity-0 graviton}},  {\em Physical Review D}
  {\bf 83} (2011), no.~10 103516,
  [\href{http://xxx.lanl.gov/abs/1010.1780}{{\tt arXiv:1010.1780}}].

\bibitem{2005IJMPA..20.2294R}
S.~{Reynaud} and M.-T. {Jaekel}, {\it {Testing the Newton Law at Long
  Distances}},  {\em International Journal of Modern Physics A} {\bf 20} (2005)
  2294--2303, [\href{http://xxx.lanl.gov/abs/gr-qc/050}{{\tt gr-qc/050}}].

\bibitem{2012PhR...513....1C}
T.~{Clifton}, P.~G. {Ferreira}, A.~{Padilla}, and C.~{Skordis}, {\it {Modified
  gravity and cosmology}},  {\em Physics Reports} {\bf 513} (2012) 1--189,
  [\href{http://xxx.lanl.gov/abs/1106.2476}{{\tt arXiv:1106.2476}}].

\bibitem{2003Natur.425..374B}
B.~{Bertotti}, L.~{Iess}, and P.~{Tortora}, {\it {A test of general relativity
  using radio links with the Cassini spacecraft}},  {\em Nature} {\bf 425}
  (2003) 374--376.

\end{thebibliography}\endgroup

\end{document}